\documentstyle[12pt]{article}
\date{}
\begin{document}

\topmargin -1cm
\textheight 9in
\oddsidemargin -1cm
\evensidemargin -1cm
\textwidth 7in
\newcommand{\nc}[2]{\newcommand{#1}{#2}}
\newcommand{\ncx}[3]{\newcommand{#1}[#2]{#3}}
\nc{\pr}{\prime}
\nc{\nl}{\newline}
\nc{\np}{\newpage}
\nc{\nit}{\noindent}
\nc{\be}{\begin{equation}}
\nc{\ee}{\end{equation}}
\nc{\ba}{\begin{array}}
\nc{\ea}{\end{array}}
\nc{\bea}{\begin{eqnarray}}
\nc{\eea}{\end{eqnarray}}
\nc{\nb}{\nonumber}
\nc{\dsp}{\displaystyle}
\nc{\bit}{\bibitem}
\nc{\ct}{\cite}
\ncx{\dd}{2}{\frac{\partial #1}{\partial #2}}
\ncx{\ddf}{2}{\frac{\delta #1}{\delta #2}}
\nc{\pl}{\partial}
\nc{\dg}{\dagger}
\nc{\cD}{{\cal D}}
\nc{\cS}{{\cal S}}
\nc{\cL}{{\cal L}}
\nc{\ag}{\alpha}
\nc{\bg}{\beta}
\nc{\gam}{\gamma}
\nc{\Gam}{\Gamma}
\nc{\bgm}{\bar{\gam}}
\nc{\del}{\delta}
\nc{\Del}{\Delta}
\nc{\eps}{\epsilon}
\nc{\ve}{\varepsilon}
\nc{\zg}{\zeta}
\nc{\th}{\theta}
\nc{\vt}{\vartheta}
\nc{\Th}{\Theta}
\nc{\kg}{\kappa}
\nc{\lb}{\lambda}
\nc{\Lb}{\Lambda}
\nc{\ng}{\eta}
\nc{\ps}{\psi}
\nc{\Ps}{\Psi}
\nc{\sg}{\sigma}
\nc{\Sg}{\Sigma}
\nc{\rg}{\rho}
\nc{\fg}{\phi}
\nc{\Fg}{\Phi}
\nc{\vf}{\varphi}
\nc{\og}{\omega}
\nc{\Og}{\Omega}
\nc{\tr}{\mbox{Tr}}
\nc{\lh}{\left(}
\nc{\rh}{\right)}
\nc{\lbr}{\left[}
\nc{\rbr}{\right]}
\nc{\lll}{\left\{}
\nc{\rll}{\right\}}
\nc{\ld}{\left.}
\nc{\rd}{\right.}
\nc{\rl}{\right|}
\nc{\bor}{\overline}
\ncx{\abs}{1}{\left| #1 \right|}
\nc{\vs}{\vspace{2ex}}
\nc{\ev}{\equiv}
\nc{\la}{\leftarrow}
\nc{\ra}{\rightarrow}
\nc{\lra}{\leftrightarrow}
\nc{\lan}{\langle}
\nc{\ran}{\rangle}
\nc{\bc}{\begin{center}}
\nc{\ec}{\end{center}}
\nc{\e}{\mbox{e}}

\title{The Faddeev-Popov term reviewed}
\author{D.E. Jaramillo$^{a,b}$, J.H. Mu\~noz$^{a,c}$, and A. Zepeda$^a$\\
$^a${\small \em Departamento de F\'\i sica, Cinvestav}, \\
{ \small\em Apartado Postal 14-740, 07000},\\
{\small \em  M\'exico D.F., M\'exico}.\\
$^b${\small \em Departamento de F\'\i sica, Universidad de Antioquia},\\
{\small \em A.A. 1226, Medell\'\i n, Colombia}.\\ 
$^c${\small \em Departamento de F\'\i sica, Universidad del Tolima},\\
{\small \em A.A. 546, Ibagu\'e, Colombia}.}
\maketitle

{\small ABSTRACT. Some textbooks and reports claim that the Jacobian
$\Del_f[A]$
which
arises in the  discussion of the Faddeev-Popov method to quantize
non-Abelian gauge
theories and which is given by the derivative of the gauge fixing
conditions over the gauge group parameters, is gauge invariant. Other
references however 
prove the opposite. In this brief report we present a discussion about
this
matter.\\}

{\small
 RESUMEN.
Algunos textos mencionan que el Jacobiano $\Del_f[A]$, el cual surge en la
discusi\'on del m\'etodo de Faddeev-Popov  para cuantizar teorias de norma
no abelianas y est\'a dado por la derivada de las condiciones que fijan la
norma con respecto a los par\'ametros del grupo de simetr\'{\i}a, es
invariante
de
norma. Otras referencias muestran lo contrario. En este trabajo se
presenta una discusi\'on sobre este hecho.}\\

PACS: 11.10.-z; 02.20.Sv; 02.30.Cj  
\newpage

Already thirty years ago L.D. Faddeev and V.N. Popov introduced their
prescription\ct{fp} to 
quantize non-Abelian gauge theories, according to which the gauge fixing
conditions give rise to a  system 
of anticommutating scalar ghost fields which enter only as internal lines
in Feynman loops.\\

In non-Abelian   gauge theories, considering only the gauge bosons,
the vacuum-to-vacuum amplitude
$\lan 0,+\infty|0,-\infty\ran\ev _+\lan 0|0\ran_-$  
is expressed by the 	functional integration \ct{feyn}
\be\label{00}
_+\lan 0|0\ran_- 
\sim\int\cD A^\mu 
\e^{iS[A^\mu ]},
\ee
where $\cD A^{\mu}=\prod_{a,x}dA_{a}^{\mu}(x)$ and the action
$\cS\ev\int\!
\mbox{d}^4x\,\cL$ are invariant 
under the gauge transformation

\be
A_\mu\ra A_\mu^\th=U^\dg A_\mu U + iU^\dg(\pl_\mu U)
\ee
with $U\ev\e^{i\th}$ and  $\th\ev\th_a T_a$ (setting the coupling 
constant equal to one). The generators $T_a$ of the simmetry group  
satisfy the algebra
\be
[T_a,T_b]=iC_{abc} T_c  .
\ee

An inmediate problem arises because of the divergent nature of the
functional integration(\ref{00}), which is due to  the 
gauge invariance of the action. Hence an  infinity factor should be 
factorized and removed before implementing the perturbative expansion.
The trick designed by Faddeev-Popov, for this purpose, begins with  the
introduction of the  Jacobian

\be\label{jaco}
\Del_f[A_{\mu}]
=\lh\int\!\cD \th \,\del\lbr f[A_\mu^\th ]\rbr\rh^{-1}
\ee
with $\cD \th=\prod_{a,x}d\th_{a}(x)$, so that we can  write  the
expression
(\ref{00}) as

\be\label{001} _+\lan 0|0
\ran_- \sim\int\!\cD A_\mu 
\,\e^{i\cS[A_{\mu}]}\,
\Del_f[A_{\mu}]
\int\!\cD \th \,\del\lbr f[A_\mu^\th ]\rbr.
\ee

$f[A^\th_\mu (x)]=0$ is called the gauge fixing condition and it should
have a
solution 
 $\th(x)$ for a given $A_\mu$ \ct{lee}. If $\th_n$ is a zero of $f[A_\mu
^\th]$, we obtain
\be
\Del_f[A]=\abs{\ddf{f}{\th}}_{\th_n}.
\ee

Some textbooks and reports \ct{noinv} claim that the Jacobian $\Del_f[A]$ 
is gauge invariant when they are explaining the quantization of
non-Abelian gauge theories. Some other references \ct{noshow} 
state without  proof that this
determinant is gauge invariant. The argument of references \ct{noinv}
about the gauge invariance  of $\Del_f[A]$ goes as follows:

\bea
\Delta_f^{-1}\left[ A_{g} \right]&=&\int \cD g^{'}\del \left( F\left[
A_{g^{'}g} \right] \right)\nb\\[3mm]
&=&\int \cD \left( g^{'}g \right) \del \left( F \left[ A_{g^{'}g} \right]
\right)\nb\\[3mm]
&=&\int \cD (g^{''}) \del \left( F \left[ A_{g^{''}} \right]
\right)
=\Del_f^{-1}[A] .
\eea

Although the last three equalities are correct, the first one is wrong
since it assumes  that the group measure $\cD g$ is equal to the
parameter measure (which enters in the definition given in Eq.
(\ref{jaco})),

\[
\cD g=\prod_{a,x} d\th_{a}(x)=\cD  \th .  
\]

In this note we show that this Jacobian is not gauge
invariant, and we give an example. At the end we explain how this
result is in agreement with references \ct{inv}.\\

After the gauge transformation $A\ra A^{-\th}$ the Jacobian 
$\Del_f[A]$
defined in the Eq.(\ref{jaco}) transforms as

\bea
\Del_f^{-1} [A^{-\th}]&=&
\int\!\cD\th''\,\del \lbr F[(A_\mu^{-\th})^{\th''}]\rbr\nb\\[5mm]
&=&	\int\!\cD\th'\,\abs{\ddf{\th''}{\th'}}
\del \lbr f[A_\mu^{\th'}]\rbr\nb\\[5mm]
\label{djaco} 
&=&\Del_f^{-1}[A]\abs{\ddf{\th'' }{\th' }}_{\th'=\th_n},
\eea
where $\th$, $\th'$ and $\th''$ are related by
\[
\e^{i\th''}=\e^{i\th}\e^{i\th'}\ev\e^{i\th \# i\th'}
\]
and the 
exponent $x\#y$ is given by an infinite Baker-Campbell-Hausdorff series
of multiple commutators \ct{haus}

\be
x\#y=x+y+\frac{1}{2}[x,y]+\frac{1}{12}\Big(\big[x,[x,y]\big]+
\big[y,[y,x]\big]\Big)+...
\ee
Therefore the variation $i\del \th''$, with respect to $\th'$
is given by

\bea
i\del\th''&=&i\th\#i(\th'+\del\th')-i\th\#i\th' \nb\\[3mm]
&=&i\del\th'-\frac{1}{2}[\th,\del\th']-\frac{i}{12}
\big[\th,[\th,\del\th']\big]+...,
\eea
so that in terms of the components of $\th$ we can write
\be
\ddf{\th''_a}{\th'_b}=\del_{ab}+\frac{1}{2}C_{abc}\th_c
+\frac{1}{12}C_{ace}C_{dbe}\th_c\th_d+...
\ee
For example in SU(2) we have  $C_{abc}=	\eps_{abc}$,

\be
\abs{\ddf{\th''_a}{\th'_b}}_{\th'_n}=1+\frac{1}{144}\left( \th_1^2 +
\th^2_2 + \th^2_3 \right) \left[ \th_1^2 + \th^2_2 + \th_3^2 + 12 \right]
+ ...
\ee
and obviously the Jacobian $\Del_f[A]$  is not 
gauge invariant.\\

Note that to get Eq. (\ref{djaco}) we have integrated over all
parameters of the
simmetry group
instead of over all group elements. We would like to stress that the
references \ct{noinv} get Eq. (\ref{djaco}) without
 the determinant $\abs{\del\th''/\del\th'}$. They
have
integrated over all group elements.\\

Now we show that we can obtain the result (\ref{djaco}) from Eq. (37)
of Zaidi's paper \ct{inv}. First, let us explain how he obtains this
equation.\\

Let us consider the functional integral $\int F[q]\cD q$ and suppose
 that we want to change the function $q(x)$ in the
functional integral by another function $q'(x)$ given by

\[
q(x)=\int K(x,y)q'(y)dy
\] 

with $K(x,y)=K(y,x).$\\

Now we wish to find out what happens in the functional integral. For this
aim we must seek the relationship between the two measures $\cD q$ and
$\cD q'$. If we expand $q(x)$ and $q'(x)$ in terms of an orthonormal
set of functions $\{ \phi (x)\}$, we obtain:

\[
\cD q =\prod_{i=1}^{\infty}dq_i=\left| \frac{\partial
q_i}{\partial q_{j}'} \right| \cD q'
\]

hence 

\be \label{zaidi}
\int F[q][dq]=\left| \frac{\partial q_{i}}{\partial q_{j}^{'}}
\right| \int F[Kq^{'}][dq^{'}]
\ee

Eq. (\ref{zaidi}) is Eq. (37) in Zaidi's paper. $q_{i}$ and
$q_{j}^{'}$ are the coefficients in the expansion of $q(x)$ and
$q^{'}(x)$, respectively.\\

If we consider $\th^{''}(x)$ and $\th^{'}(x)$ instead of $q(x)$ and
$q^{'}(x)$, respectively, and expand them in terms of the set of
generators $\{ T_a \}$ of the gauge transformation as:

\bea\nb
\th^{''}(x)=\sum_{a} \th_{a}^{''} (x) T_{a}, && \th^{'}(x)=\sum_{b}
\th_{b}^{'}(x) T_{b},
\eea
with $F[q]=\Del _{f}[A]$, we obtain Eq. (\ref{djaco}) from
Eq. (\ref{zaidi}). In this case $\th^{''}(x)$ and $\th^{'}(x)$ are related
by 
$\e^{i\th''}=\e^{i\th}\e^{i\th'}\ev\e^{i\th \# i\th'}$. Also we can see
that the  equivalent expression to $K(x,y)$ is not symmetric.\\

We also can obtain the result (\ref{djaco}) from Eq. (15.5.17) of
Weinberg's book \ct{inv}. First, let us mention that Eq. (15.5.1) of
this reference,

\[
 I=\int \cD \phi  g[\phi] B\left[ f[\phi] \right] \left| F[\phi] \right|,
\]
is equal to Eq. (\ref{001}) with the following correspondence:

\bea\nb
g[\phi]&=&\e^{iS[A^\mu ]},\\ B[f[\phi]]&=&\del \left[ F[A_{\mu}]
\right],\nb \\ \left| F[\phi] \right|&=&\Del_{f}[A] \nb,
\eea
where the F-matrix is

\be
F_{\alpha x, \beta y}[\phi]=\left.\frac{\del
f_{\alpha}[\phi_{\lambda};x]}{\del
\lambda_{\beta}(y)} \right|_{\lambda =0}.
\ee

If we consider the gauge transformation with parameters
$\rho^{\alpha}(x; \Lambda , \lambda)$ in the $\phi$ fields as the product
of the gauge transformation with parameters
 $\Lambda^{\alpha}(x)$  followed   by the
gauge transformation with parameters $\lambda^{\alpha}(x)$, we obtain

\be
F_{\alpha x, \beta y}[\phi_{\Lambda}]=\int J_{\alpha x, \gamma z}[\phi
, \Lambda] R^{\gamma z}_{\beta y}[\Lambda]d^{4}z ,
\ee
with
\begin{eqnarray*}
J[\phi ,\Lambda]=\left.\frac{\partial f_{\alpha}[\phi_{\rho};x]}{\partial
\rho^{\gamma}(z)}\right|_{\rho =\Lambda}, &&
R^{\gamma z}_{\beta y}=\left.\frac{\partial \rho^{\gamma}(z; \Lambda
,\lambda)}{\partial \lambda^{\beta}(y)}\right|_{\lambda =0},
\end{eqnarray*}
hence

\be\label{wein} 
\left| F [\phi_{\Lambda}] \right|=\left| J[\phi ,\Lambda] \right|
\left|
R[\Lambda]\right| .
\ee

Eq. (\ref{wein}) is Eq. (15.5.17) in  Weinberg's book, and it 
is
equal to Eq. (\ref{djaco}). Weinberg has introduced a
weight-function $\rho (\Lambda)$ as

\be
\rho (\Lambda)=1/\left| R[\Lambda] \right|,
\ee
thus, $\rho (\Lambda)$ is $\abs{\del\th''/\del\th'}^{-1}$.\\

Finally, if we use the gauge invariance of the action
$\cS$ and the measure $\cD 
A^\mu$ combined with (\ref{djaco}), 
after the gauge
transformation $A\rightarrow A^{-\th}$,
the integral (\ref{001}) can be written as

\be\label{00a} _+\lan 0|0
\ran_- \sim\int\!\cD A^\mu \,\e^{i\cS[A_{\mu}]}\,
\,\del\Big[ f[A_\mu ]\Big] \Del_f[A_{\mu}]\int\!\cD \th 
\abs{\ddf{\th''}{\th'}}_{\th'=\th_n}^{-1},
\ee
so that removing the entire factor 
$\dsp{\int\!\cD \th\abs{\frac{\del \th''}{\del\th'}}^{-1}}$ 
we write (\ref{00a}) as
\be\label{002}
_+\lan 0|0 \ran_-
\sim \int\!\cD A_\mu \,
\del\Big[ f[A_\mu ]\Big]\,\Del_f[A_{\mu}]\,\e^{i\cS[A_{\mu}]}.
\ee
which is the Faddeev-Popov prescription.\\

We notice that the contribution of  $\Del_f[A]$ is contained
in the infinity factor which is removed from the
integration. Thus, the expression (\ref{002}) is obtained
independently of whether
$\Del_f[A]$ is gauge invariant or not.  If $\Del_f[A]$
were gauge invariant, then one would remove just $\int \cD \th$.\\

In conclusion, we have showed that the Jacobian $\Del_{f}[A]$ is not gauge
invariant and given an example. Also we have explained how this result
can be obtained from references \ct{inv}.\\ 

\vspace{1cm}
{\bf Acknowledgements:}	 This work was partially supported by
COLCIENCIAS(Colombia) and CONACYT(M\'exico).

\end{document}